\documentclass[11pt,cites,graphicx]{article}
\usepackage{epsfig, times, graphics, amsmath, amssymb}

\title{Nonequilibrium Thermodynamics of Time-dependent Langevin Systems:\\ Energy Balance Relation and the Extended Form of the Second Law}

\author{Hao Ge\footnote{Mathematical Department
and Center for Computational Systems Biology, Fudan University,
Shanghai 200433, P.R. China; Email: gehao@fudan.edu.cn}}

\begin{document}
\maketitle
\renewcommand{\thefootnote}{\fnsymbol{footnote}}

\begin{abstract}
A general nonequilibrium thermodynamic theory is developed for
time-dependent Langevin dynamics, starting from the common
definition of nonequilibrium Gibbs entropy. It is shown that the
notations appearing in the First and the Second Law of
thermodynamics could be consistently applied to transient
nonequilibrium processes. We find out a general equality
representing the energy balance relation for the system, and further
an explicit mathematical interpretation for the extended form of the
Second Law is proposed, which only comes into existence in the
time-dependent case. More applications to several classic
thermodynamic processes are finally included.
\end{abstract}

Many different kinds of approach have been put forward in the last
several decades \cite{Lebon2008}, but in contrast to equilibrium
systems, with their elegant theoretical framework, the understanding
of nonequilibrium steady-state systems is still primitive. In 1998,
Oono and Paniconi \cite{OP} proposed a framework of steady state
thermodynamics, and distinguished the steadily generated heat which
is generated even when the system remains in a single steady state
and the total heat. They call the former the ``housekeeping heat'',
which is equal to the entropy production in steady state and may
come from the chemical driven force in biochemical systems
\cite{QH05,QHBe05}. Subtracting the housekeeping heat from the total
heat defines the excess heat, which reflects the time-dependent
variation of the system. Moreover, they also put forward a
phenomenological extended form of the Second Law: ``A process
converting work into excess heat is irreversible. And
`reversibility' is modulo house-keeping heat, which is produced
anyway''.

On the other hand, if one needs to study the nonequilibrium
thermodynamics from a microscopic or mesoscopic point of view, and
wants to be less ambitious and get rather satisfactory
understanding, effective stochastic models maybe the best approach
to choose. And in recent years it has been realized that a
trajectory perspective of stochastic processes might encode
surprisingly more information than one might expect from traditional
thermodynamic arguments \cite{Jar2008,Seifert08}. Inspired by Oono
and Paniconi's framework \cite{OP}, Hatano and Sasa \cite{HS}
derived the first explicit expression for the extended form of the
Second Law of Thermodynamics between steady states, namely
$T\triangle S\geq Q_{ex}$, where $S$ is the general entropy defined
in their paper, and $Q_{ex}$ is the excess heat.

Therefore, one can actually use Markov chains and diffusion
processes as mathematical tools to study nonequilibrium states. It
could help us break through the shackles of former equilibrium and
near-equilibrium statistical mechanics, and would accomplish a
rather self-contained theory of nonequilibrium thermodynamics, which
will be discussed in the present article.

Here we consider the dynamics of a Brownian particle in a circuit
driven by an external force, i.e.

$$\gamma \dot{X}(t)=-\frac{\partial V(x;t)}{\partial x}|_{x=X(t)}+f(X(t),t)+\xi(t),$$
where $\xi(t)$ represents Gaussian white noise whose intensity is
$2\gamma kT$ according to the Einstein's relation. We employ
periodic conditions as Kurchan and Hatano-Sasa have done in their
previous works \cite{Kur,HS,Seifert05}. This time-dependent system
is realized by changing the time-dependent potential $V(x,t)$ and
nonconservative force $f$.

Langevin differential equations which govern a random variable X can
also be reformulated as ``Fokker-Planck'' differential equations
(Fokker-Planck Equations), or ``master equations'', which govern the
probability distribution $p(x,t)$ of $X(t)$. Denote the drift
coefficient $b(x,t)=\frac{-\frac{\partial V(x;t)}{\partial
x}+f(x,t)}{\gamma}$, it says
\begin{equation}\label{FP_Diff}
\frac{\partial p(x,t)}{\partial t}=-\frac{\partial j(x,t)}{\partial
x},
\end{equation}
where the current
$j(x,t)=b(x,t)p_t(x)-\frac{kT}{\gamma}\frac{\partial
p(x,t)}{\partial x}$.

We write the steady-state probability distribution function as
$\pi(x,t)$ for which the right-side of Eq. (\ref{FP_Diff}) vanishes
for any fixed $t$, i.e.
$$\frac{\partial [b(x,t)\pi(x,t)-\frac{kT}{\gamma}\frac{\partial \pi(x,t)}{\partial x}]}{\partial x}=0.$$

First of all, we accept the opinion that for the Second Law, in
particular, a proper formulation and interpretation of entropy is
more subtle. For stochastic processes, the so-called Gibbs entropy
has already been widely accepted \cite{Groot62,Jaynes65}, in
statistical physics as well as information theory, hence it becomes
our starting point.

Denote the real distribution of $X(t)$ at time $t$ as $\{p(x,t)\}$,
and we define the {\em general entropy} at time $t$ as
$$S(t)=-k\int
p(x,t)\log p(x,t)dx.$$

It is widely known that the entropy change $dS$ could be to
distinguished in  two terms \cite{NP,QH06,QH01a}: the first, $d_eS$
is the transfer of entropy across the boundaries of the system, and
the second $d_iS$ is the entropy produced within the system.

Here, it is easy to derive that \cite{QH01a,QH06}
\begin{equation}\label{EntropyEq}
\frac{dS(t)}{dt}=d_iS+d_eS=epr(t)-hdr(t),
\end{equation}
where $epr(t)=d_iS=\int \frac{\gamma j(x,t)^2}{Tp(x,t)}dx\geqslant
0$ is the entropy production rate at time $t$, and
$hdr(t)=d_eS=\frac{\gamma}{T}\int b(x,t)j(x,t)dx$ is due to the
exchange of heat with the exterior, called the heat dissipation
rate.

Consequently, although individual particle has no knowledge of
ensemble properties, or even whether there is an ensemble, U.
Seifert successfully developed a theory of entropy production along
the trajectory \cite{Seifert05}, which was suggested by the
so-called fluctuation theorem \cite{ECM,Jar1,Cro2}.

He defined the sample trajectory entropy along the trajectory at
time $t$ as
$$S(X(t),t)=-k\log p(X(t),t),$$
recalling that $p(x,t)$ is the real distribution of the Langevin
dynamics $\{X(t)\}$ at time $t$.

The equation for the motion of $S(X(t),t)$ \cite[Eq.9]{Seifert05}
becomes
$$\frac{d S(X(t),t)}{dt}=epr(X(t),t)-hdr(X(t),t),$$
where
$epr(X(t),t)=-k\frac{\partial_tp(x,t)}{p(x,t)}|_{x=X(t)}+\frac{\gamma
j(x,t)}{Tp(x,t)}|_{x=X(t)}\dot{X}(t)$ is the sample entropy
production at time $t$ in which $\dot{X}(t)$ is of the Stratonovitch
type, and $hdr(X(t),t)=\gamma\frac{b(x,t)}{T}|_{x=X(t)}\dot{X}(t)$
is the sample heat dissipation in the medium.

The rational after these identifications for the change rate of
entropy becomes clear when averaging over all the trajectories. It
gets $epr(t)=\langle epr(X(t),t)\rangle,$ and $hdr(t)=\langle
hdr(X(t),t)\rangle$, due to an heuristic equality $\langle
\dot{X}|x,t\rangle=\frac{j(x,t)}{p(x,t)}$ \cite{Seifert05}.

The sample heat dissipation $hdr(X(t),t)$ could be regarded as the
total heat conduction $Q_{tot}(X(t),t)$ with the medium, i.e.
$$Q_{tot}(X(t),t)=T\cdot hdr(X(t),t)=\gamma b(X(t),t)\dot{X}(t),$$
 and its ensemble
average $Q_{tot}(t)=T\cdot hdr(t)$. By convention, we take the sign
of heat to be positive when it flows from the system to the heat
bath.

The idea of decomposing the total heat into a ``housekeeping'' part
and another ``excess'' part was put forward by Oono and Paniconi
\cite{OP}, and made explicit in Langevin systems by Hatano and Sasa
\cite{HS}.

Now we could define the other two kinds of heat: the housekeeping
heat and excess heat along the trajectory
\cite{HS,GQ_JMP2007,GJ_JSP2008}:
$$Q_{hk}(X,t)=\gamma[b(x,t)-\frac{kT}{\gamma}\frac{\partial \log\pi(x,t)}{\partial x}]|_{x=X(t)}\dot{X}(t),$$
$$Q_{ex}(X,t)=kT\frac{\partial \log\pi(x,t)}{\partial x}|_{x=X(t)}\dot{X}(t),$$
and obviously $Q_{tot}(X,t)=Q_{ex}(X,t)+Q_{hk}(X,t)=\gamma
b(X(t),t)\dot{X}(t)$.

After averaging over all the trajectories, we found out that the
housekeeping heat is always nonnegative, which implies the
nonequilibrium essence of the system:

\begin{eqnarray}
Q_{hk}(t)&=&\int \gamma(b(x,t)-\frac{kT}{\gamma}\frac{\partial \log\pi(x,t)}{\partial x})j(x,t)dx\nonumber\\
&=&\int \gamma(b(x,t)-\frac{kT}{\gamma}\frac{\partial
\log\pi(x,t)}{\partial x})^2p(x,t)dx\geq 0.\nonumber
\end{eqnarray}

For equilibrium system, $Q_{ex}$ reduces to the total heat
$Q_{tot}$, because in this case $Q_{hk}\equiv 0$ due to
$b(x,t)=\frac{kT}{\gamma}\frac{\partial \log\pi(x,t)}{\partial x}$.
And in time-independent steady state, $Q_{ex}(t)\equiv 0$, hence the
housekeeping heat $Q_{hk}$ equals the work done by the external
driven force, which is all dissipated \cite{QH05,QH06,QHBe05}.

However, the situation is quite different for the time-dependent
nonequilibrium system, in which the housekeeping heat still comes
from the work done by some external driven force but where does the
excess heat come from?We will show that its natual origin is just
the change of a thermodynamic quantity called {\em general internal
energy} in the form of heat.

On the other hand,
\begin{eqnarray}
Q_{ex}(t)&=&\langle Q_{ex}(X,t)\rangle\nonumber\\
&=&kT\int \frac{\partial \log\pi(x,t)}{\partial x}j(x,t)dx\nonumber\\
&=&-kT\int \frac{\partial j(x,t)}{\partial x}\log\pi(x,t)dx\nonumber\\
&=&\frac{\partial p(x,t)}{\partial t}\log \pi(x,t)dx,\nonumber
\end{eqnarray}
due to the Fokker-Planck equation (\ref{FP_Diff}).

To derive the energy balance condition and introduce the concept of
{\em ``general internal energy''}, we shall accept the definition of
``the dissipative work'' done on the system \cite{HuS,GJ_JSP2008}:
$$W(X(t),t)=-kT\frac{\partial \log\pi(x,t)}{\partial t}|_{x=X(t)},$$
and its ensemble average becomes $W(t)=\langle
W(X(t),t)\rangle=-kT\int p(x,t)\frac{\partial \pi(x,t)}{\partial
t}dx$.

If the system satisfies the detailed balance conditions for all
time, i.e. $b(x,t)=\frac{kT}{\gamma}\frac{\partial
\log\pi(x,t)}{\partial x}$, then the traditional concept of internal
energy exists and both of the excess heat and dissipative work
contribute to its change, which is the First Law undoubtedly
\cite{GQ_JMP2007}. Therefore, we believe that the situation will not
be essentially different even if detailed balance conditions fails.

Notice that the integral of the dissipative work subtracting the
excess heat does not depend on the particular ``path'' taking
through the parameter space, namely, only depends upon the initial
and final states. Thus, there exactly exists a {\em ``general
internal energy''} $U(t)=-kT\int p(x,t)\log\pi(x,t)dx$, whose
derivative is just the difference of the dissipative work and excess
heat, i.e.
\begin{equation}\label{Firstlaw}
\frac{dU(t)}{dt}=-Q_{ex}(t)+W(t).
\end{equation}

For the equilibrium canonical ensemble, it is just the ordinary
internal energy according to the Maxwell-Boltzmann's law. Hence here
Eq.(\ref{Firstlaw}) is just the generalized First Law of
thermodynamics. Also from the trajectory view, we can define the
internal energy along the trajectory $X(t)$:
$U(X(t),t)=-kT\log\pi(X(t),t)$. Then one has $\langle
U(X(t),t)\rangle=U(t)$ and $U(X(t),t)=-Q_{ex}(X(t),t)+W(X(t),t)$,
which implies the First Law is also satisfied along every
trajectory.

Nevertheless, the situation becomes more complicated in
nonequilibrium case, since there may exists some external driven
force which also pumps energy into the system but does not
contribute to the change of internal energy \cite{QHBe05}.

Denote the work done by the external driven force as $Edf(t)$, and
we here try to figure out its relationship with other quantities
defined previously.

Now we understand that there exist totally two kinds of external
works done on the system, one is the dissipative work $W(t)$ and the
other $Edf(t)$ from the external driven force. They result in the
change of general internal energy and the heat dissipation
respectively. Hence according to energy balance, one has

$$W(t)+Edf(t)=\frac{dU(t)}{dt}+T\cdot hdr(t).$$

On the other hand, we have already known that

$$\frac{dU(t)}{dt}=-Q_{ex}(t)+W(t)=-Q_{tot}(t)+Q_{hk}(t)+W(t),$$
and also $Q_{tot}(t)=T\cdot hdr(t)$.

Therefore, it yields $Edf(t)=Q_{hk}(t)\geq$, which is only known to
be valid in steady state \cite{QH05,QHBe05} before.

Now, we turn to the Second Law. Although all the thermodynamic
quantities in the previous sections could be defined along the
sample trajectory, the Clausius inequality and many other
thermodynamic constrains related to the Second Law should be
interpreted statistically through ensemble average.

Based on the elementary definition of free energy in equilibrium
thermodynamics $F=U-TS$, here we could define a {\em general free
energy} in the same way:
$$F(t)=U(t)-TS(t)=kT\int p(x,t)\log\frac{p(x,t)}{\pi(x,t)}.$$

For equilibrium system, it is just the Gibbs free energy in a
spontaneously occurring chemical reaction at constant pressure $p$
and temperature $T$, and also the Helmholtz free energy for systems
at constant $V$ and $T$ \cite{Ross2008}. Its change gives the
maximum work, other than pV work. Therefore, it is called a ``hybrid
free energy'' by Ross, J. \cite{Ross2008}.

From a mathematical point of view, it is just the relative entropy
of the distribution $\{p(x,t)\}$ with respect to another one
$\{\pi(x,t)\}$. Hong Qian \cite{QH01a} has proved that this relative
entropy from information theory could be identified as the free
energy difference associated with a fluctuating density in
equilibrium, and is also associated with the distribution deviate
from the equilibrium sate in nonequilibrium relaxation.

Then,
\begin{eqnarray}\label{FreeEq}
\frac{dF(t)}{dt}&=&\frac{dU(t)}{dt}-T\frac{dS(t)}{dt}\nonumber\\
&=&W(t)-Q_{ex}(t)-T(epr(t)+hdr(t))\nonumber\\
&=&W(t)-(T\cdot epr(t)-Q_{hk}(t)),
\end{eqnarray}

Here we introduce a new concept named {\em Free heat} $Q_f(t)=T\cdot
epr(t)-Q_{hk}(t)$ identifying the free energy change in the form of
heat, i.e.
$$\frac{dF(t)}{dt}=W(t)-Q_f(t),$$
which will play the central role in the extended form of the Second
Law of Thermodynamics.

 Notice that the entropy production rate
$epr(t)=d_iS=\int \frac{\gamma j(x,t)^2}{Tp_t(x)}dx$ is nonnegative,
hence according to Eqs. (\ref{EntropyEq}) and (\ref{FreeEq}), we
derived several thermodynamic inequalities in the differential form:

\begin{subequations}
\label{Secondlaw}
\begin{eqnarray}
T\frac{dS(t)}{dt}+Q_{tot}(t)&=&T\cdot epr(t)\geq 0,\label{Secondlaw1}\\
\frac{dF(t)}{dt}-W(t)-Q_{hk}(t)&=&-T\cdot epr(t)\leq
0.\label{Secondlaw2}
\end{eqnarray}
\end{subequations}

Eq. (\ref{Secondlaw1}) is just the well-known Clausius inequality
($dS\geq -\frac{Q_{tot}}{T}$), which is rectified to obtain
expressions for the entropy produced ($dS$) as the result of heat
exchanges ($Q_{tot}$). And Eq. (\ref{Secondlaw2}) is a general
version of the free energy inequality for the amount of work
performed on the system, since the work values must then be
consistent with the Kelvin-Planck statement \cite{Finn93} and
forbids the systematic conversion of heat to work.

More precise, the quantity $Q_{hk}(t)$ in Eq. (\ref{Secondlaw2})
vanishes when the detailed balance condition holds£¬ and then it
returns back to the traditional Helmhotz or Gibbs free energy
inequalities of equilibrium thermodynamics depending on whether it
is a NVT or NPT system \cite{BQ2008}. In this case, $-dF\geq -W$,
which implies the decrease of free energy gives the maximum
dissipative work done upon the external environment.

Also, their corresponding integral forms are
\begin{subequations}
\label{Secondlawint}
\begin{eqnarray}
T\triangle S+\int Q_{tot}(t)dt&\geq& 0,\label{Secondlawint1}\\
\triangle F-\int W(t)dt-\int Q_{hk}(t)dt&\leq&
0.\label{Secondlawint2}
\end{eqnarray}
\end{subequations}

Beyond the traditional form of the Second law based on entropy
production, we would give a rather different but much more general
expression that what Hatano and Sasa got \cite{HS}, only need to
notice that the free heat
$$Q_f(t)=T\cdot epr(t)-Q_{hk}(t)=(kT)^2\int p(\frac{\partial \log\frac{\pi(x,t)}{p(x,t)}}{\partial x})^2dx\geq 0.$$

Then according to Eqs. (\ref{EntropyEq}) and (\ref{FreeEq}), another
group of  inequalities in the differential form arises:

\begin{subequations}
\label{Secondlaw_ex}
\begin{eqnarray}
T\frac{dS(t)}{dt}+Q_{ex}(t)&=&T\cdot epr(t)-Q_{hk}(t)\geq 0,\label{Secondlaw_ex1}\\
\frac{dF(t)}{dt}-W(t)&=&-Q_f(t)=-T\cdot epr(t)+Q_{hk}(t)\leq 0.
\label{Secondlaw_ex2}
\end{eqnarray}
\end{subequations}
followed by their corresponding integral forms
\begin{subequations}
\label{Secondlawint_ex}
\begin{eqnarray}
T\triangle S+\int Q_{ex}(t)dt&\geq& 0,\label{Secondlawint_ex1}\\
\triangle F-\int W(t)dt&\leq& 0.\label{Secondlawint_ex2}
\end{eqnarray}
\end{subequations}

Eq. (\ref{Secondlawint_ex1}) is the extended form of Clausius
inequality during any nonequilibrium time-dependent process, whose
special case is included in Hatano and Sasa's work \cite{HS}. And
Eq. (\ref{Secondlawint_ex2}) is a different general form of free
energy inequality. It implies the dissipative work value must be
consistent with the Oono-Paniconi statement of the extended Second
Law of thermodynamics \cite{OP}, which forbids the systematic
conversion of excess heat to work.

For equilibrium case, $Q_{ex}=Q_{tot}$, then they both return back
to Eq. (\ref{Secondlaw1}) actually. And then if in nonequilibrium
steady state, then $Q_f(t)\equiv 0$, and this extended form of the
Second Law is concealed.

Finally, we apply the previous formula to several important
thermodynamic processes:

First, the relaxation process towards steady states. The relaxation
process towards steady states has been extensively discussed by
Glansdorff and Prigogine \cite{GlPr,NP}, and then by Schnakenberg
for the master-equation systems \cite{Sc}.

Here, the First Law of Thermodynamics becomes
\begin{equation}
\frac{dU(t)}{dt}=-Q_{ex}(t),\nonumber
\end{equation}
and the energy balance reads
\begin{equation}
\frac{dU(t)}{dt}+T\cdot hdr(t)=Edf(t).\nonumber
\end{equation}

Combined with $Edf(t)=Q_{hk}(t)$, it is found that the whole heat
dissipation is from two sources: one is the excess heat contributing
to the change of general internal energy; the other is the
housekeeping heat caused by the external driven force.

Furthermore, the extended form of the Second Law now gives
\begin{equation}
\frac{dF(t)}{dt}=-T\cdot epr(t)+Q_{hk}(t)=-Q_f(t)\leq 0,\nonumber
\end{equation}
thus $F(t)$ serves as a Lyapunov function for this relaxation
process, which actually has a solid thermodynamic basis.

Second, the thermodynamic cyclic process. In equilibrium
thermodynamics, a thermodynamic cycle is a series of thermodynamic
processes which returns a system to its initial state. As a
conclusion of cyclic process, all the state variables should have
the same value as they had at the beginning. Thus $\triangle
U=\triangle S=\triangle F=0$.

But variables such as heat and work are not zero over a cycle, but
rather are process dependent. The First Law of Thermodynamics
dictates that the net heat input is equal to the net work output
over any cycle, i.e. $\int W(t)dt=\int Q_{ex}(t)dt$.

Hence in this case, the former form of the Second Law
(\ref{Secondlawint}) gives

\begin{subequations}
\label{CyclicSecondlaw}
\begin{eqnarray}
\int Q_{tot}(t)dt=T\cdot \int epr(t)dt&\geq& 0,\label{CyclicSecondlaw1}\\
\int W(t)dt&\geq& -\int Q_{hk}dt.\label{CyclicSecondlaw2}
\end{eqnarray}
\end{subequations}

If one rewrite Eq. (\ref{CyclicSecondlaw2}) as $\int
(W(t)+Edf(t))dt=\int Q_{tot}(t)dt\geq 0$, then it is just the
familiar statement of traditional Second Law of Thermodynamics ``the
conversion from work to total heat is irreversible''.

Moreover, the extended form (\ref{Secondlawint_ex}) gives
\begin{subequations}
\label{CyclicSecondlaw_ex}
\begin{eqnarray}
\int Q_{ex}(t)dt=\int Q_f(t)dt&\geq& 0,\label{CyclicSecondlaw1_ex}\\
\int W(t)dt=\int Q_{ex}(t)dt&\geq& 0,\label{CyclicSecondlaw2_ex}
\end{eqnarray}
\end{subequations}
which explicitly confirms the claim that ``the conversion from work
to excess heat is irreversible'' \cite{OP}. In other words, during a
cyclic process, not only the total heat but also the excess heat
could only be from the system into the heat bath rather than follow
the opposite direction.

Third, the transition process between two steady states. The
previous steady-state thermodynamics of Langevin systems is based on
a generalized version of the Jarzynski work relation
\cite{Jar1,Jar2,Jar2008}, and concluded that $Q_{ex}$ should
correspond to the change of a generalized entropy $S$ in an
appropriate limit \cite{HS}. Actually, the extended form of the
Second Law derived by Hatano and Sasa \cite{HS} is just a
straightforward consequence of (\ref{Secondlawint_ex}), which
satisfied for any arbitrary transient state.

It is indispensable to point out that the general entropy defined in
their work \cite{HS} is just the general internal energy related to
the First Law of Thermodynamics rather than the general Gibbs
entropy in the present article and also in Seifert's recent work
\cite{Seifert05}. Note that the two quantities are always different
except for steady states, hence our formula is consistent with that
of Hatano and Sasa.

Summarily, for stochastic systems, the central problem is around the
extension of the Second Law, which originally describes the
fundamental limitation on possible transitions between equilibrium
states. And recently, it has been studied from the trajectory point
of view, which stimulated the rapid emergence of so-called
fluctuation theorems \cite{Seifert08,Jar2008}. Thus the main purpose
of the present article is to investigate the relationship and
rationality of the thermodynamic functionals along the trajectory,
especially the present formula here. It would be interesting to test
experimentally all the quantities and relations, especially in
nonharmonic time-dependent potentials, where one does not expect
Gaussian distributions. Although we only study the stochastic
process, the thermodynamic relations derived here may be satisfied
much more generally, especially the extended form of the Second Law,
since both of the entropy production rate and housekeeping heat
could be gained phenomenological \cite{OP}.

\section*{Acknowledgment}

The author is very grateful to Prof. Min Qian and Prof. Hong Qian
for helpful discussion.

\end{document}